\documentclass{article}

\usepackage{arxiv}

\usepackage[utf8]{inputenc}
\usepackage[T1]{fontenc}    
\usepackage{url}           
\usepackage{booktabs}       
\usepackage{amsfonts}      
\usepackage{nicefrac}       
\usepackage{microtype}      
\usepackage{lipsum}		
\usepackage{graphicx}
\usepackage{doi}
\usepackage{derivative}
\usepackage{placeins}
\usepackage[ddmmyyyy]{datetime}
\usepackage{xcolor}
\usepackage{nicematrix}
\usepackage{caption}
\usepackage{multirow}
\usepackage{float}
\floatstyle{plaintop}
\restylefloat{table}
\usepackage{longtable}
\usepackage{lscape}
\usepackage{makecell}
\usepackage{siunitx}
\usepackage{array}
\usepackage{vcell}
\usepackage{comment}
\usepackage{amsmath}
\usepackage{amssymb}
\usepackage{hyperref}
\usepackage[numbers, sort]{natbib}
\newcolumntype{L}[1]{>{\raggedright\let\newline\\\arraybackslash\hspace{0pt}}p{#1}}
\newcolumntype{C}[1]{>{\centering\let\newline\\\arraybackslash\hspace{0pt}}p{#1}}
\newcolumntype{R}[1]{>{\raggedleft\let\newline\\\arraybackslash\hspace{0pt}}p{#1}}

\newcommand{\ra}[1]{\renewcommand{\arraystretch}{#1}}

\title{Efficient Multi-Change Point Analysis to decode Economic Crisis Information from the S\&{}P500 Mean Market Correlation}

\newcommand*\samethanks[1][\value{footnote}]{\footnotemark[#1]}

\author{\href{https://orcid.org/0000-0003-0529-7926}{\includegraphics[scale=0.06]{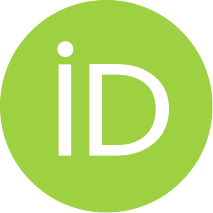}\hspace{1mm}Martin ~Heßler}\thanks{Center for Nonlinear Science, University of Münster, 48149 Münster, Germany} \\
	Institute for Theoretical Physics\\
	University of Münster\\
	48149 Münster, North Rhine-Westphalia, Germany\\
	\texttt{m\_{}hess23@uni-muenster.de} \\
	\And
	\href{https://orcid.org/0000-0001-9391-1601}{\includegraphics[scale=0.06]{orcid.pdf}\hspace{1mm}Tobias ~Wand}\samethanks\\
	Institute for Theoretical Physics\\
	University of Münster\\
	48149 Münster, North Rhine-Westphalia, Germany\\
	\texttt{t\_{}wand01@uni-muenster.de}\\
 \And
	\href{https://orcid.org/0000-0003-0986-0878}{\includegraphics[scale=0.06]{orcid.pdf}\hspace{1mm}Oliver ~Kamps} \\
	Center for Nonlinear Science\\
	University of Münster\\ 
	48149 Münster, North Rhine-Westphalia, Germany\\
	\texttt{okamp@uni-muenster.de} \\
}
\renewcommand{\headeright}{\emph{arXiv} Preprint}
\renewcommand{\undertitle}{\emph{arXiv} Preprint}
\renewcommand{\shorttitle}{Efficient Change Point Analysis to decode Economic Crisis Information}

\hypersetup{
pdftitle={Efficient Multi-Change Point Analysis to decode Economic Crisis Information from the S\&500 Mean Market Correlation},
pdfsubject={physics.data-an, q-fin.ST},
pdfauthor={Martin Hessler, Tobias Wand, Oliver Kamps},
pdfkeywords={Bayesian Multi-Change Point Analysis, Linear Trend Segment Fit, Computationally Efficient Open-Source Python Implementation, S\&P500, Mean Market Correlation, Economic Crises, Econophysics},
}

\begin{document}
\maketitle

\begin{abstract}
Identifying macroeconomic events that are responsible for dramatic changes of economy is of particular relevance to understand the overall economic dynamics. We introduce an open-source available efficient Python implementation of a Bayesian multi-trend change point analysis which solves significant memory and computing time limitations to extract crisis information from a correlation metric. Therefore, we focus on the recently investigated \textit{S\&P500} mean market correlation in a period of roughly 20 years that includes the dot-com bubble, the global financial crisis and the Euro crisis. The analysis is performed two-fold: first, in retrospect on the whole dataset and second, in an on-line adaptive manner in pre-crisis segments. The on-line sensitivity horizon is roughly determined to be 80 up to 100 trading days after a crisis onset. A detailed comparison to global economic events supports the interpretation of the mean market correlation as an informative macroeconomic measure by a rather good agreement of change point distributions and major crisis events. Furthermore, the results hint to the importance of the U.S. housing bubble as trigger of the global financial crisis, provide new evidence for the general reasoning of locally (meta)stable economic states and could work as a comparative impact rating of specific economic events.
\end{abstract}


\section{Introduction}
Previous work \cite{Stepanov2015,RandomMatrixLalouxPotters,RandomMatrixStanley} has identified the mean correlation of the S\&P500's stock time series to capture essential features of the global market movement by uncovering a strong agreement between the mean correlations, the maximum eigenvalue of the correlation matrices and the first principal components of the stock correlations.
Following this observation, the one dimensional time series should contain valuable information about economic dynamics, such as e.g.  bubbles and economic crises. But if the information is stored in the condensed mean correlations, we should consequently be able to extract that information in some way from the strongly fluctuating time series. A demonstration of that would first, lead to new insights into how the information is encoded in the mean correlation and second, consolidate --- independently from previous works --- the significant role of the mean correlation as a global market observable. In this article, we find that the quantitatively most probable points of changing mean correlation trends identified by a Bayesian change point (CP) analysis show rather good correspondence to the major economic crisis events of the observed time period, i.e. the Dot-com bubble, the financial and the Euro crisis. A related approach of a CP analysis is used in a publication of Dehning et al. \cite{a:dehning2020} to quantify the efficiency of government decisions during the COVID-19 pandemics. Nevertheless, there are significant differences to our research task: First, the authors of \cite{a:dehning2020} have access to well-established pandemic models, e.g. the susceptible-infected-recovered (SIR) model, whereas we have to assume linear segments with different slopes in absence of better models for the mean correlation dynamics. Second, Dehning et al. can determine a certain range for change point positions based on known government actions and incorporate gradual changes of parameters directly in the model equations of the Monte Carlo runs. In contrast, we determine the CP probability distributions with respect to all possible CP configurations based on the given data without employing Monte Carlo methods.\\
We demonstrate that the approach lends itself for extracting information about economic state history, could help to separate endogenous from exogenous crises and implies that it might be possible in a way to derive precursors for endogenous events from the mean correlation. Naturally, the approach applied in this article cannot be interpreted in terms of precursor information, since it is applied off-line and in retrospect to the \textit{whole} mean correlation time series, i.e. it contains data from before and after the major economic events. This holds also if we change into an on-line perspective in which we successively add future data and perform the CP analysis on each updated time series. The CP will only manifest itself if we include a minimum amount of data that encodes the uprising trend change. However, the on-line sensitivity horizon of the method to detect trend changes in the investigated time series lies roughly between $80$ and $100$ trading days which might provide a chance to identify current events for experts to focus on with respect to possible new and locally quasi-stationary economic states. Depending on the time horizon of macroeconomic research it might even provide some evidence for ongoing changes of the market dynamics.\\
Nonetheless, the separation of the investigated time period into segments of almost constant linear trends offers a rather straight-forward interpretation in terms of temporally stable market periods. Even if it is not possible to interpret these states as abstract conceptional entities, as it would be possible e.g. by the clustering approach that is described in Stepanov et al. \cite{Stepanov2015}, the method allows for a more direct evaluation of the data and a closer connection to real events. It is also robust with respect to intrinsic parameter choices which is not guaranteed for clustering algorithms and the pre-assumed number of clusters.\\   
The remainder of this article is structured as follows: Section \ref{sec:Methods} explains the data gathering and preprocessing in \ref{sec:Data} and the Bayesian methodology used for the change point detection in \ref{sec:ChangePointsMethod}. Section \ref{sec:Results} shows the ex post identified change points in \ref{sec:ChangePointResults} and an on-line analysis in \ref{sec:OnlineChangePointResults}. The results are compared in detail to a time-line of global economic events of the period. Finally, we discuss and compare our results to the work of Stepanov et al. \cite{Stepanov2015} in section \ref{sec:Discussion}.

\section{Data and Methods}
\label{sec:Methods}
The preprocessing of the data that we analyse is shortly summarized in subsection \ref{sec:Data}. More details about the CP analysis method can be found in subsection \ref{sec:ChangePointsMethod}.
\subsection{Data Preparation}
\label{sec:Data}
The chosen procedure to calculate the mean market correlation of the \textit{S\&P500} stock index is chosen analogously to the articles \cite{Stepanov2015,MemoryEffectsSP500}. Confined to the considered time period between 1 January 1992 and 28 December 2012 we filter for companies that are included in the \textit{S\&P500} index for at least $\SI{99.5}{\percent}$ of the time and interpolate occasionally missing data. The data is used to compute pairwise correlations $C_{i,j}$ of the locally normalised returns of the remaining 249 daily-resolved stock price time series for 5291 trading days. The correlations are computed on overlapping rolling data windows of size $\tau=42$. Shown by the principle component analysis in \cite{Stepanov2015}, the mean correlation
\begin{equation}
    \Bar{C} = \frac{1}{N} \sum_\textmd{i,j} C_{i,j}
\end{equation}
is in high agreement with the first principle component. In addition, it describes the greatest part of the variability in the data. Thus, we focus on the mean market correlation $\Bar{C}(t)$ in which the mean correlation of each window is identified with its 21st time stamp. The corresponding time series is shown in blue in figure \ref{fig:correlation data}. More details about the preprocessing procedure can be found in Appendix \ref{Appendix A}.
\begin{figure}
    \centering
    \includegraphics[width = \textwidth]{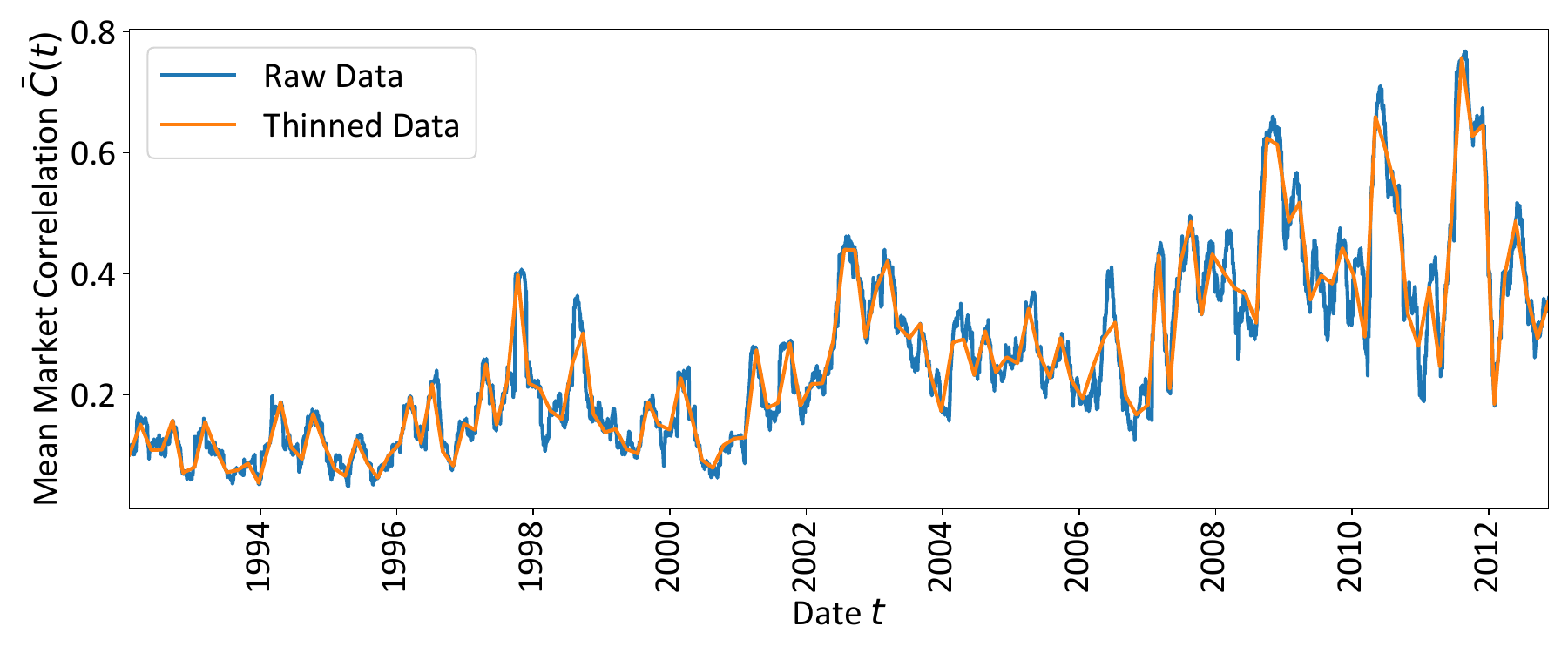}
    \caption{The calculated mean correlation of the \textit{S\&P500} market. The time series are the mean values of correlation matrices which were calculated based on moving windows of length $\tau=42$ days plotted against the 21st time stamp of the $\tau$-days-interval. The original raw correlation time series is computed on a sequence of data windows that are successively shifted by one trading day. The thinned version which contains every 40th data point of the raw time series, i.e. 132 data points in total, is used to avoid exorbitant computation times due to the combinatorics of CP configurations involved in the CP analysis.}
    \label{fig:correlation data}
\end{figure}

\subsection{Change Point Identification}
\label{sec:ChangePointsMethod}
Since the mean market correlation time series $\Bar{C}(t)$ should contain valuable information about changing market characteristics, we investigate its changing trends over a time period of stable and crises states in detail. Therefore, we use a tool that considers the change points (CPs) of linear trends over time $t$. The open-source Python package \textit{antiCPy} \cite{MartinHessler, url:GitHessler2021, url:DocsHessler2021} implements the Bayesian analysis of changing linear trends, introduced by von der Linden et al. \cite{dose2004bayesian, von2014bayesian}. By assuming $m$ CPs in the trend, the time series data $\Bar{C}(t)$ is modelled by linear segments $\phi_K$ of different slopes, formally given by
\begin{equation}
\phi_K (t|\underline{E}, \underline{\Bar{C}}_{\rm ord}, \mathcal{I}) = \Bar{C}_K \frac{t_{K+1}-t}{t_{K+1} - t_K} + \Bar{C}_{K+1}\frac{t-t_K}{t_{K+1}-t_K},
\end{equation}  
with $t_K\leqslant t \leqslant t_{K+1}$ and the CP vector $\underline{E}$ with the entries $E_K = t_{K+1}$ for $K = 1,2,...,N-2$, whereby $N$ denotes the size of the correlation time series 
$\Bar{C}(t)$ and the vector $\underline{\Bar{C}}_{\rm ord}$ contains the design ordinates $\Bar{C}_K$ of the assumed CP positions, i.e. the positions where  consecutive segments are connected with each other. Note that the first and last allowed position of a CP is restricted to the second and penultimate time stamp $t$. For each possible configuration $\underline{E}$ of CPs the corresponding posterior probability $p(\underline{E}|\Bar{C}(t),\underline{t},\mathcal{I})$ is calculated via Bayes' theorem
\begin{equation}
    p(\underline{E}|\Bar{C}(t),\underline{t},\mathcal{I}) = \frac{p(\underline{E} \mid \underline{t}, \mathcal{I}) \cdot p(\Bar{C}(t) \mid \underline{t}, \underline{E}, \mathcal{I})}{p(\Bar{C}(t) \mid \underline{t}, \mathcal{I})}  \label{eq:changePointPosterior}
\end{equation}
with the underlying time steps $\underline{t}$ and the background information $\mathcal{I}$. Following this reasoning, we can derive the most probable CP configurations quantitatively if we perform the calculations for all feasible CP combinations, i.e. a number of
    \begin{equation}
        Z_m = \binom{N-2}{m} = \frac{(N-2)!}{m!((N-2)-m)!}
    \end{equation}
CP configurations and select the most probable ones. Furthermore, the complete pdf of CP configurations enables us to calculate the most probable linear segment fit and confidence bands for each time step by calculating the averaged sum of the linear segment fits at that time weighted with their respective probabilities. Doing so, even future values and their confidence bands can be extrapolated based on the current data information in a formally consistent manner. In addition, the marginal CP probability density functions can be computed by averaging the joint probabilities at each time stamp $t$ with respect to its ordinal position in each configuration.\\
However, performing these calculations proves to be almost prohibitive (cf. also \cite{von2014bayesian}) if we recall the combinatorial number of CP configurations, e.g. for the whole time series $\Bar{C}(t)$ of $5269$ data points for only three CPs, i.e.
    \begin{equation}
        Z_2 = \binom{5267}{3} \approx 2.4\cdot 10^{10} 
    \end{equation}
combinations. One idea to solve this issue for long time series with moderate number of change points is presented by von der Linden et al. \cite{von2014bayesian}: The approach uses a Monte Carlo sum over a limited number of random CP configurations to estimate the linear segments. However, with increasing number of change points even a Monte Carlo sum approach is not feasible anymore, since the extremely high-dimensional CP configuration space is sampled too sparsely to derive reliable fit results. Furthermore, the time to compute the linear segment fits serially explodes up to several decades or hundreds of years. For this reason, first we decide to thin the mean market correlation data $\Bar{C}(t)$ by storing only every 40th data point and performing the calculations on this dataset of $132$ points which keep the trend characteristics of the data very well as visible in figure \ref{fig:correlation data}. Nevertheless, a naive serial implementation of the CP analysis would fail with regards to the computation time and --- maybe even worse --- collapses due to memory limitations which do not allow for storing all possible change point configurations at the same time. Imagine e.g. the situation of the thinned time series under the assumption of five CPs: We easily run into problems if we want to store a matrix of dimensions $(286.243.776 \times 7)$ with double precision. To avoid such issues, we take advantage of the efficient implementation of the change point analysis in the \textit{antiCPy} package which involves parallelisation and an algorithm that constructs on the fly and savely organised only a suitable subset of CP configurations for each parallel worker while avoiding any redundancy which would compromise the results. More details of the practical implementation will be prepared in a follow-up manuscript.
\section{Results}
\label{sec:Results}
In subsection \ref{sec:ChangePointResults} we apply the CP analysis to the whole mean market correlation dataset $\Bar{C}(t)$ under the assumption of two up to four CPs. The results are double-checked by analyses with one and five CPs as documented in Appendix \ref{Appendix B}. In addition, we perform an on-line adaptive CP analysis in the three pre-crisis segments in subsection \ref{sec:OnlineChangePointResults} to complete the analysis.
\subsection{Change Point Analysis}\label{sec:ChangePointResults}
\begin{figure}
    \centering
    \includegraphics[width = 0.95\textwidth]{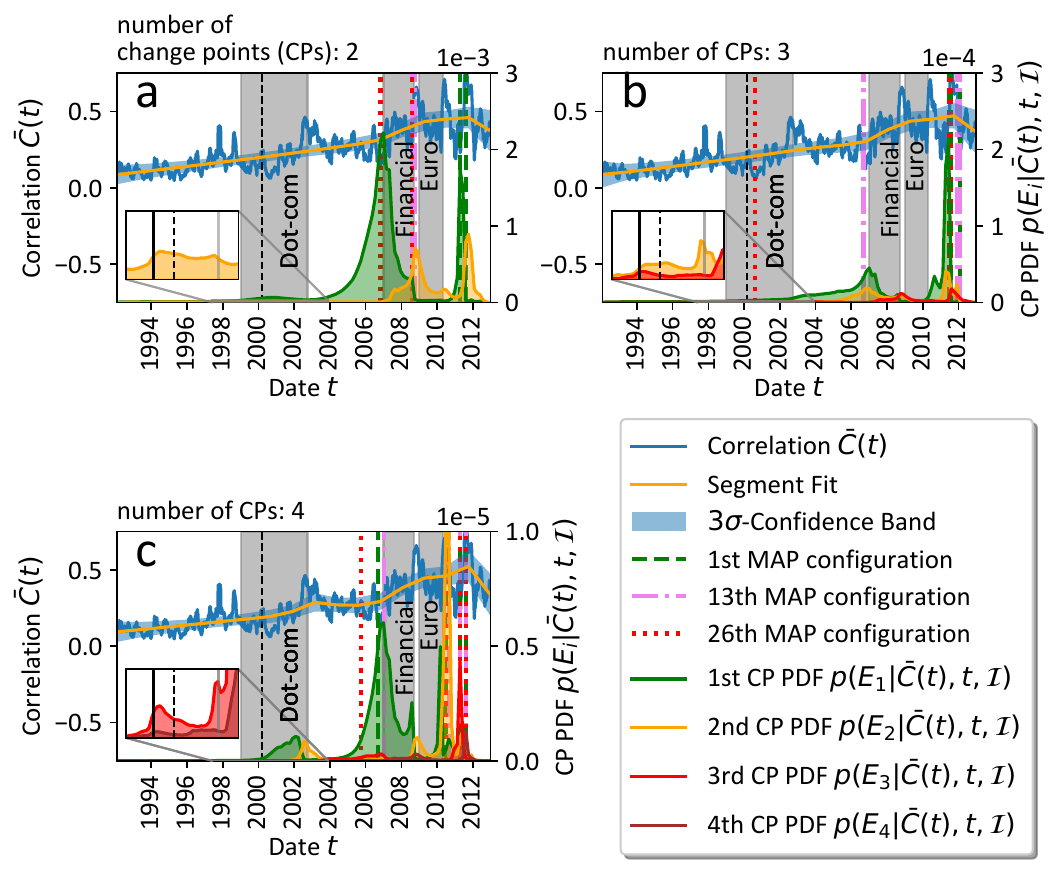}
    \caption{The change point (CP) analysis results under the assumption of two, three and four CPs is shown in (a), (b) and (c), respectively. The $x$-axis ticks correspond to the 1 January of each year. The major economic events as the dot-com bubble, the global financial crisis and the Euro sovereign debt crisis are indicated by the grey shaded areas: The dot-com bubble's rise starts around 1 January 1999, reaches its all time high at the dashed vertical line and ends around the 3 October 2022 with a after- bubble trough. The Financial crisis starts with the burst of the U.S. housing bubble around 1 January 2007 and culminates with the Lehman Brothers' bankruptcy on 15 September 2008 which marks the end of gray shaded area. The Euro crisis starts with the first insolvencies of European banks starting with the 1 January 2009 and culminates in the downgrading of Greece's sovereign debt rating on 27 April 2010. The shown CP probability densities (PDFs) exhibit consistently peaks in the areas of major economic disturbance. Furthermore, three of the most probable CP configurations are indicated by vertical lines as an example consistently with the CP PDFs and the signed crisis events. The insets zoom into some peaks of the higher order CP PDFs to illustrate that their profile also contains valuable information, e.g. for the dot-com bubble in (a) and (b). At least the first CP PDF in (a) is slightly increased shortly after the dot-com bubble's all time high, but in (b) only the inset provides insights into the dot-com bubble period. The linear segment fit with corresponding confidence intervals is the PDF-weighted average of the linear segment fits at that time and is shown in orange to illustrate the agreement between (a-c). For a detailed discussion of the results we refer to the running text. Additional results for the assumption of one and five CPs can be found in Appendix \ref{Appendix B}, figure \ref{fig:SI CP}.}
    \label{fig:CPanalysis}
\end{figure}
As already mentioned above the one-dimensional mean market correlation $\Bar{C}(t)$ is expected to contain valuable information about the global market dynamics including major crisis events. However, extracting this information in a reliable manner is an ambitious task. Based on our assumption that the linear trends of the market correlation encode some information about changing dynamics due to crisis events, we perform a detailed CP analysis as described in subsection \ref{sec:ChangePointsMethod}. We compute the probability density functions (PDFs) of the CPs assuming one up to five trend-CPs hidden in the data to confirm that our results are stable against these a priori assumptions. Even the results based on only one assumed CP, which obviously cannot reflect the complicated trend structure in the data, are roughly consistent with the more elaborate fits and the same is true for models with more than four CPs. The interested reader can take a look at the results for one and five CPs in Appendix \ref{Appendix B}, figure \ref{fig:SI CP}.\\
The most suitable assumptions should contain at least three CPs, since the investigated period from 31 January 1992 until 28 December 2012 contains three important major economic events: the dot-com bubble, the financial and the the Euro crisis. Therefore, in figure \ref{fig:CPanalysis} of the main article we show our results from two up to four CPs in increasing order and compare them to these three major economic events. Unfortunately, the continuously progressing dynamics of economy can make it rather difficult to determine exact dates of crisis onset and decline.\\
In table \ref{table: economic events} we provide a timeline of selected significant events of the three major economic bubbles and crises and compare them to the time stamps at which maxima occur in the CP PDFs of figure \ref{fig:CPanalysis}. Some maxima, especially of the higher order CP PDFs are only visible by appropriate up-scaling. For example the high peak of the first CP PDF in \ref{fig:CPanalysis}(c) is accompanied by a very small one in (a). In figure \ref{fig:CPanalysis}(b) it is not found.
\subsubsection{Dot-Com Bubble}
First, let us consider the dot-com bubble: We highlight in grey approximately the time interval in which the \textit{NASDAQ Composite Index} starts to grow relatively fast from 1 January 1999 up to 1 October 2002, the day at which the \textit{NASDAQ Composite Index} fell roughly to its lowest value after the bubble's all-time high marked by the black dotted vertical line on 3 October 2000. The trough of 1 October 2002 is accompanied by a more general stock market downturn in 2002 which started around the 11 September 2001 terrorist attacks \cite{o:911}. The most interesting observation for the dot-com bubble is that around its onset and decline the marginal CP PDFs of the second (cf. figure \ref{fig:CPanalysis}(a)), second and third (b) or third and fourth (c) CP are pronounced as visible in the insets. For orientation in the insets, the beginning and end of the dot-com interval are indicated by black and grey solid vertical lines, respectively. In figure \ref{fig:CPanalysis}(a) and (c) the CP PDFs of the first (a) and first and second CP (c), respectively, also reach local maxima shortly after the bubble's all-time high. These are plausible results, since the dot-com bubble was of speculative nature which should become visible in the studied correlations of the \textit{S\&P500} stock index returns without notably delay. For completeness, we mention the 14 May 1999 as an important economic date independent from the dot-com bubble. On this day a sharp rise in consumer prices in the U.S. led to a drop in the stock market \cite{o:consumer_price_rise_stock_drop}. Even if the width of the CP PDFs and their pronounced profile at the end of the bubble tend to support mostly a connection with the continuous strong rise of the \textit{NASDAQ Composite Index}, of course our approach cannot decide definitely whether events like the 14 May 1999 or the aftermath of the Asian financial crisis \cite{o:asian_crisis} around 1997-1998 or Russian financial crisis \cite{o:russian_crisis} around 1998-1999 also might influence the CP PDF profiles. Keeping that in mind, the correspondence of CP positions and onset/decline of the dot-com bubble nevertheless strengthens the hypothesis that the mean market correlations trends encode distinct market periods.
\subsubsection{Financial Crisis}
Second, we consider the financial crisis. While the dot-com bubble was mostly created by speculative investment decisions, less speculatively induced economic crises like the global financial crisis include a complex cascade of sub-crisis events and cannot be simplified in that manner. Therefore, we proceed with indicating their approximate beginning up to their culmination events.
The very pronounced peaks of the first (a), the first and second  (b) and the first and third (c) CP PDFs are located rather exactly in the time interval from 31 October 2006 up to 3 January 2007 which coincides with the burst of the U.S. housing bubble, the trigger event for the global financial crisis of the later years \cite{HousingBubbleBurst}. Nevertheless, the commonly accepted definition of the beginning of the financial crisis is on 9 August 2007 when the interest rates for inter-bank financial loans rose sharply \cite{o:common_financial_crisis_onset,o:common_financial_crisis_onset2}. This emphasises the problem of determining exact dates for crisis events. Our results suggest that the U.S. housing bubble is clearly notable as a change in the mean market correlation, whereas the change in interest rates for inter-bank financial loans is not detected by our CP analysis. This is a rather expected result, since we consider stock return correlations: The correlations probably capture financial events like strong price changes of stock bubbles rather accurately, but their response to real-world events (like the changing interest rates) might be delayed.\\  
However, to demonstrate the precise coincidence we mark the burst of the U.S. housing bubble around 1 January 2007 as the beginning of the grey shaded area for the global financial crisis and end with the Lehman Brothers' insolvency on 15 September 2008 \cite{Wiggins2014}.
Also for the Lehman Brothers' bankruptcy at least the first and second CP PDFs exhibit local maxima regardless of assumed number of CPs. Note that similar to the dot-com bubble's key dates there is almost no delay between the event and the detected trend change, since the Lehman Brothers' bankruptcy had an almost immediate effect on the world economy. Only two weeks later on 29 September 2008 the U.S. stock market collapsed \cite{BailoutReject}. 
\subsubsection{Euro Crisis}
Third, we discuss the Euro crisis as the direct consequence of the global financial crisis. We indicate its earliest start when the first ten European banks asked for a bailout up to January 2009, represented in the figure by the 1 January 2009, and end with the culmination of the crisis when Greece's sovereign debt rating was downgraded on 27 April 2010 by \textit{Standard \& Poor's} \cite{o:culminating_Greece_downgrade}. The temporal proximity and strong causal connection of the global financial crisis and the Euro crisis are captured by the width of the local maxima of the second (a), third (b) and first, second and fourth (c) CP PDFs in the passage from one crisis to the other. Also the commonly mentioned culmination event of the downgrading of Greece's sovereign debt rating is mirrored by the CP analysis results. Even if the downgrading alone might have only a moderate influence on the mean market correlation trends in terms of symbolic manifestation of the crisis, the indicated culmination event is mirrored by small peaks in the second and first CP PDF in (a) and (b), respectively, and in the well-pronounced peaks of the first and second CP PDFs in (c).\\
Apart from these considerations, the CP PDFs in the cases (a-c) show pronounced maxima in the second half of 2010 up to December 2011. It seems to be impossible to deduce origin events for each individual CP PDF peak, since the time period is full of economic (and political) events including a series of sovereign debt downgradings of European countries, bailout plans, the Arab Spring and more, summarised in table \ref{table: economic events}. But in spite of that it seems to be rather probable that the trend changes detected by the above mentioned CP PDFs are correlated with the turbulent economic situations at that time.
\subsubsection{Remarks and Intermediate Conclusion}
Since the CP PDFs are computed over the thinned grid of time steps only every 40th trading day is sampled. For the definition of the maxima we examine these really calculated support points. In cases of modes of a certain width we yield intervals of higher CP PDF. Of course, the listed events are not complete and its content should not be considered as causal deduction between the listed real world events and the CP PDF peaks. It should rather illustrate that the uncovered times of high trend change PDFs in fact correlate with highly turbulent times in the economy. The maxima of the CP PDFs are clustered around events that are connected with the dot-com bubble, the global financial crisis and the Euro sovereign debt crisis.\\
All in all, the results support the hypothesis that the trend in the mean market correlation encodes valuable information about major economic events that can be uncovered by applying a Bayesian trend CP analysis. As the similarities of the time stamps for CP PDF peaks in table \ref{table: economic events} illustrate, the results are comparable to a high extent, regardless of how many change points we assume. For an extended robustness check compare the results in figure \ref{fig:SI CP} under the assumption of one and five CPs in Appendix \ref{Appendix B}. It might be unsurprising that sharp drops and bursting bubbles like the dot-com bubble or the U.S. housing bubble are often well-captured by our approach, since we rely on correlations of stock returns which show these changes almost immediately. Nevertheless, considering the mean market correlation with a CP analysis enables us to eliminate trend changes that manifest only in a small subset of stock prices and thus, acts like a filter to focus only on economic events of global importance. Furthermore, in some cases the ex post derived CP PDF peaks could help to identify important economic events or even rate the impact of specific time periods or isolated economic events. Moreover, the general ansatz can be adapted to other economic observables or subsets of stock prices, e.g. national companies, companies of certain sectors or of certain size etc., to filter for economic events that are of special interest for the considered subset. However, we emphasise that the analysis is performed on the whole available data which makes it an ex post analysis that cannot be interpreted in terms of an ex ante economic risk predictor.\\
We conclude the CP analysis discussion with some remarks on the consistency of the results. The blue mean market correlation $\Bar{C}(t)$ in figure \ref{fig:CPanalysis} is well-fitted by the orange segment fits with corresponding $3\sigma$-credibility bands (CBs). The credibility bands quantify the certainty of the given fit based on the surrounding data information. In that sense the narrow shape of the CBs suggests that the fit performs well for the underlying mean market correlation data $\Bar{C}(t)$. A comparison of the very similarly shaped fits (a-c) also confirms the robustness of the results. Finally, we studied the individual most probable CP configurations three of which are drawn by green dashed, violet dash-dotted and red dotted vertical lines in figure \ref{fig:CPanalysis} as an example. In (a-c) the most probable configurations are clustered around the CP PDF peaks of the global financial crisis and the Euro sovereign debt crisis. The dot-com bubble is included only a bit later in the most probable configurations, but consistent with the less pronounced CP PDFs of that period. However, the first configuration that includes a date between the 1 January 1999 and the 1 October 2002 is identified to be the 264th of $8385$ total configurations in (a), the 26th combination of $357760$ in (b) and the 354th  of $11358880$ configurations in (c). This corresponds to the most probable $\SI{3.15}{\percent}$, $\SI{7.28}{\cdot10^{-5}\percent}$ and $\SI{3.12}{\cdot10^{-5}\percent}$ configurations in (a),(b) and (c), respectively, and round out our analysis.
\begin{landscape}
\ra{1.5}
\begin{longtable}{L{3cm} L{5cm} C{1cm}R{3.2cm}R{3.2cm}R{3.2cm}}
   \toprule[1.3pt]
    \multicolumn{1}{l}{\textbf{Event Date}} & \multicolumn{1}{l}{\textbf{Event Description}} & \multicolumn{1}{c}{\textbf{Source}}& \multicolumn{3}{r}{\textbf{CP PDF Peak Dates}}\\
    \cmidrule(l{1.75em}){4-6}
    &&&  \textbf{Figure} \ref{fig:CPanalysis}\textbf{(a)} & \textbf{(b)} & \textbf{(c)}\\ \cmidrule{1-6} \endhead
    1994 & Over the year bond prices fell continuously in consequence of partially unexpected and repeated raise of federal funds rates by the \textit{FED}. In consequence bonds lost about \$1.5 trillion in market value globally. & \cite{o:bond_massacre} & 2 February 1995 to\newline 26 July 1995;\newline 31 March 1995 to\newline 21 September 1995 & 2 February 1995;\newline 31 March 1995 to\newline 21 September 1995 & 26 July 1995 to\newline 16 November 1995;\newline 2 February 1995;\newline 26 July 1995 to\newline 16 November 1995\\
    December 1994 & Due to devaluation of the peso by the Mexican government and thus anticipating further devaluations, investors rapidly withdrew capital from Mexican investments. In January 1995 the U.S. government coordinates a \$40 billion bailout. & \cite{o:mexican_peso_crisis} & 2 February 1995 to\newline 26 July 1995;\newline 31 March 1995 to\newline 21 September 1995 & 2 February 1995;\newline 31 March 1995 to\newline 21 September 1995 & 26 July 1995 to\newline 16 November 1995;\newline 2 February 1995;\newline 26 July 1995 to\newline 16 November 1995 \\
    1997 - 1998 & Asian financial crisis in East and Southeast Asia. Multiple origins are discussed. However, long-lasting global contagion stayed away and the markets recovered fast in 1998-1999. & \cite{o:asian_crisis} & 14 May 1999 & 10 October 1997;\newline 18 March 1999 & 10 October 1997;\newline 18 March 1999;\newline 14 May 1999 \\
    1998 - 1999 & Russian financial crisis in which the rubel was devalued. As a result many neighbouring countries experienced also severe crises. & \cite{o:russian_crisis} & 14 May 1999 & 18 March 1999 to\newline 13 July 1999 & 18 March 1999;\newline 14 May 1999\\
    1 January 1999 & The modest onset of the dotcom-bubble 1998 turns into a rapid rally of the \textit{NASDAQ Composite Index}. In the literature the long pre-bubble period begins commonly in 1995. & \cite{o:dotcom, o:dotcom_data} & 14 May 1999 & 18 March 1999 to\newline 12 July 1999 & 18 March 1999\\
    14 May 1999 & Stock market drop due to sharp rise of consumer prices to \SI{1.75}{\percent} in the U.S. & \cite{o:consumer_price_rise_stock_drop} & 14 May 1999 & 14 May 1999 to\newline 12 July 1999 & 14 May 1999 \\
    10 March 2000 & Based on the \textit{NASDAQ Composite Index} the Dot-com bubble reaches its all-time high. & \cite{o:dotcom_data, o:dotcom} & 16 October 2000 & 22 June 2000;\newline 9 April 2001 &kink at\newline 22 June 2000;\newline 9 April 2001\\
    11 September 2001 & In course of a terror attack three airplanes crash into the twin towers of the U.S. World Trade Center and the Pentagon coordinated by the  militant Islamist extremist network al-Qaeda. & \cite{o:911} & 24 May to\newline 23 July 2002 & 4 October 2001;\newline 24 May 2002;\newline 23 July 2002 & 4 October 2001;\newline 30 January 2002;\newline 24 May 2002;\newline 23 July 2002;\newline 18 September 2002 \\ 
    1 October 2002 & Based on the \textit{NASDAQ Composite Index} the Dot-com bubble reaches a new trough after the bubble burst in course of a more general stock downturn since 11 September 2011. & \cite{o:dotcom, o:dotcom_data} & 28 July 2005 to\newline 18 November 2005 & 28 October 2003 to\newline 24 December 2003 & 28 October 2003 to\newline 24 December 2003 \\
    2002/2003 & The Venezuelan General Strike massively hinders oil exports among others to the USA. & \cite{o:venezuela_general_strike,o:oil_energy_crisis} & 28 July 2005 to\newline 18 November 2005 & 28 October 2003 to\newline 24 December 2003 & 28 October 2003 to\newline 24 December 2003\\
    2003 & Approximate beginning of the boom-phase of the U.S. housing bubble. & \cite{a:griffin2021} & 28 July 2005 to\newline 18 November 2005 & 28 October 2003 to\newline 24 December 2003 & 28 October 2003 to\newline 24 December 2003\\
    2003 - 2016 & Begin of rising oil prices from  under \$25/bbl to above \$30/bbl. The peak of \$147.30/bbl is reached in July 2008, before they return temporarily to \$35/bbl in 2009. & \cite{o:oil_energy_crisis, o:oil_price_data} & 28 July 2005 to\newline 18 November 2005 & 28 October 2003 to\newline 24 December 2003 & 28 October 2003 to\newline 24 December 2003 \\
    Q4 2006 to Q1 2007 & Period of most intense burst of the U.S. housing bubble with new lowest price in 2011. & \cite{HousingBubbleBurst} & 2 November 2006 to\newline 3 January 2007 & 31 October 2006;\newline 2 January 2007 & 2 November 2006 to\newline 3 January 2007 \\
    9 August 2007 & Commonly listed onset of the global financial crisis. On this day the interest rates for inter-bank financial loans rose sharply. & \cite{o:common_financial_crisis_onset2,o:common_financial_crisis_onset} & 22 August 2007 & 22 August 2007 & 22 August 2007 \\
    2008 & Year of the financial crisis and great recession. The year is characterised by manifold bailouts for banks and stock market crashes. & \cite{loser2009global} & overall high CP PDFs in 2008 & overall high CP PDFs in 2008 & overall high CP PDFs in 2008\\
    January to March 2008 & The housing prices continue to collapse. The government passes a tax rebate bill on 13 February 2008 and the \textit{FED} starts bail out programs in the beginning of March. & \cite{TaxAct,o:FED_starts_bailouts} & 4 August 2008 & 4 August 2008 & 4 August 2008 \\
    30 July 2008 & U.S. governments passes bailout laws for Fanny Mae and Freddie Mac. & \cite{HousingAct} & Kink at\newline 4 August 2008 & 4 August 2008 & 4 August 2008 \\
    7 September 2008 & U.S. government's take over of Fanny Mae Association and Freddie Mac Corporation. & \cite{FannieMae} & 1 October 2008 to\newline 26 November 2008 & 1 October 2008 & 1 October 2008 to\newline 26 November 2008 \\
    15 September 2008 & Lehman Brothers bankruptcy. & \cite{Wiggins2014} & 1 October 2008 & 1 October 2008  &  1 October 2008  \\
    29 September 2008 &  Stock market collapsed when the bailout bill was rejected by the U.S. House of Representatives. & \cite{BailoutReject} & 1 October 2008 & 30 September 2008 & 1 October 2008\\
    24 October 2008 &  Many world stock indices lost around \SI{10}{\percent}. & \cite{10percentstocks} & 26 November 2008 & 26 November 2008 & 26 November 2008 \\
    1 December 2008 &  When the National Bureau of Economic Research officially declared that the U.S. was in a recession since December 2007, the \textit{S\&P500} lost \SI{8.93}{\percent} and the financial stocks of the index even \SI{17}{\percent} based on these news. & \cite{o:December2008Crash} & high CP PDF after\newline 26 November 2008 & high CP PDF after\newline 26 November 2008 & high CP PDF after\newline 26 November 2008\\
    January 2009 & The first ten European banks ask for bailout programs. & \cite{o:first_Euro_bailout_requests} & high CP PDF after\newline 26 November 2008 & high CP PDF after\newline 26 November 2008 & high CP PDF after\newline 26 November 2008 \\
    Early 2010 to mid 2012 & The Arab Spring. Great anti-government protests in large parts of the Middle East and North Africa. The oil prices rise above \$100/bbl. & \cite{o:arab_spring2, o:arab_spring, o:oil_price_data} & 5 May 2010 to\newline 1 July 2010 & 5 May 2010 to\newline 1 July 2010 & 5 May 2010 to\newline 1 July 2010 \\   
    27 April 2010 & Greece's sovereign debt rating was downgraded by \textit{Standard\& Poor's}. & \cite{o:culminating_Greece_downgrade} & 5 May 2010 to\newline 1 July 2010 & 5 May 2010 to\newline 1 July 2010 & 5 May 2010 to\newline 1 July 2010\\
    6 May 2010 & So-called flash crash led to a \SI{9}{\percent} drop in the \textit{Dow Jones Index}  caused due to high frequency trading. & \cite{o:flash_crash} & 5 May 2010 to\newline 1 July 2010 & 5 May 2010 to\newline 1 July 2010 & 5 May 2010 to\newline 1 July 2010 \\
    8 May 2010 & Passing drastic bailout plans for Greece in Brussels. A €110 billion package is approved. & \cite{o:Greece_Bailout2,Greece_Bailout} & 5 May 2010 to\newline 1 July 2010 & 5 May 2010 to\newline 1 July 2010;\newline 28 August 2010 & 5 May 2010 to\newline 1 July 2010;\newline 1 July to\newline 28 August 2010\\
    17 May 2010 & The Euro currency falls into first four-years low. & \cite{o:euro_fouryears_low} & 5 May 2010 to\newline 1 July 2010 & 5 May 2010 to\newline 1 July 2010;\newline 28 August 2010 & 5 May 2010 to\newline 1 July 2010;\newline 1 July to\newline 28 August 2010\\
    Q1/2 2010 & Several downgradings of debt ratings/bonds of European countries, i.e. Portugal, Spain, Greece. & 
    \cite{o:DowngradeSpainPortugalGreece}
    & 5 May 2010 to\newline 1 July 2010 & 5 May 2010 to\newline 1 July 2010;\newline 28 August 2010 & 5 May 2010 to\newline 1 July 2010;\newline 1 July to\newline 28 August 2010;\newline 3 September 2010 \\
     4 June 2010 & The Euro currency falls into second four-years low. Major American markets fall more than \SI{3}{\percent}. & \cite{o:second_euro_fouryears_low} & 5 May 2010 to\newline 1 July 2010 & 5 May 2010 to\newline 1 July 2010;\newline 28 August 2010 & 5 May 2010 to\newline 1 July 2010;\newline 1 July to\newline 28 August 2010\\
    November 2010 & Bailout request by Ireland. In the end of November Ireland receives €85 billion.  & \cite{Ireland_Bailout,o:Ireland_Bailout2} & 15 April 2011 & 15 April 2011 & 15 April 2011 \\
    7 April 2011 & In the evening of 6 April 2011 Portugal's government announces that it is the third after Greece's and Ireland's governments that will ask for a bailout package. On 17 May 2011 a bailout package of around €78 billion is formally adopted. & \cite{o:Portugal_Bailout} & 15 April 2011 to\newline 14 June 2011 & 15 April 2011 to\newline 14 June 2011 & 15 May 2011 to\newline 14 June 2011 \\
    13 June 2011 & Greece's credit rankings become worst in the world. & \cite{GreeceWorst} & 13 June 2011 & 13 June 2011 & 13 June 2011 \\
    August 2011 & Under the \textit{Securities Markets Programme} the ECB restarts to purchase a significant amount of eurozone sovereign bonds. Spanish and Italian yields breach \SI{6}{\percent}.
    & \cite{ECB_SMP, o:SpanishItalianYields_breach} & 6 October 2011 & 9/10 August 2011 & 10 August 2011 \\
    \bottomrule[1.3pt]\\
\caption{Overview about economic events. The table is not claimed to be complete, but serves as chronological orientation. The CP PDF peaks of figure \ref{fig:CPanalysis} are printed for easy comparison, but should be interpreted carefully, since the analysis does not allow for causal inference.}\\
\label{table: economic events}
\end{longtable}
\end{landscape}

\subsection{On-line Change Point Evolution in Pre-Crisis-Segments}
\label{sec:OnlineChangePointResults}
\begin{figure}
    \centering
    \includegraphics[width = \textwidth]{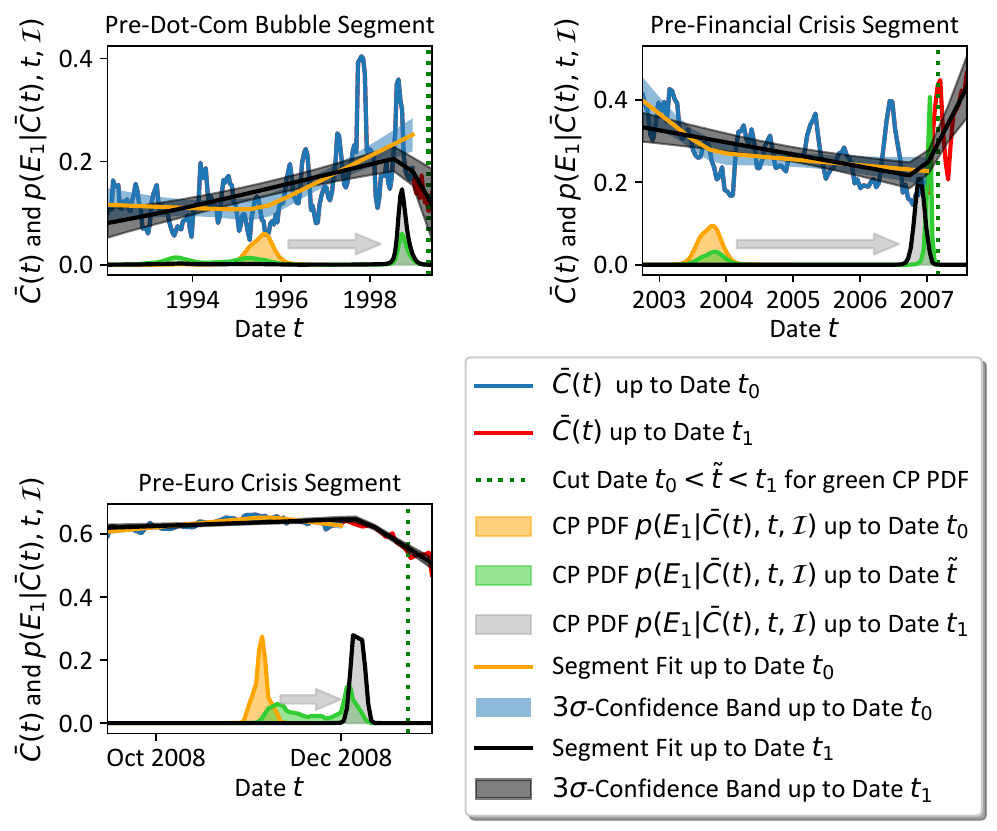}
    \caption{Results of a CP analysis on specific subsets of the whole mean market correlation $\Bar{C}(t)$ under the assumption of one CP. The blue pre-crisis segments start on 31 January 1999 (time series onset), 1 October 2002 (end dot-com bubble) and 15 September 2008 (Lehman Brothers' bankruptcy) for the dot-com bubble, the financial crisis and the Euro crisis, respectively, and end with the dates $t_0$ at the crisis onsets on 1 January 1999 (dot-com bubble), 1 January 2007 (financial crisis) and slightly before the crisis onset on 1 December 2008 (Euro crisis). The time series are thinned by using every tenth data point for the dot-com bubble and the financial crisis. The resulting orange CP PDFs are compared to the black PDFs that are computed on the red updated time series segments up to dates $t_1$, i.e. 1 June 1999 (dot-com bubble), 9 August 2007 (financial crisis) and 1 January 2009 (Euro crisis). As expected the major crisis onsets are captured by the black CP PDFs after some in-crisis data is included, whereas the orange CP PDFs may reflect consequences of economic events confined to the segment interval. Interpreted from an on-line analysis perspective, the appearance of a pronounced and narrow peak (cf. black CP PDFs) might help to identify important current economic events that may influence future economic states more than the ones detected by the orange smaller an wider CP PDFs. In contrast to the more pronounced black CP PDF in the former two cases, a higher similarity of the orange and black CP PDF is observed in the Euro crisis case. This might indicate the difficulties of isolating the Euro crisis from the financial crisis as a cascade of events in the end 2008 implies lots of trend changes with an overall negative mean market correlation trend in the end of 2008. The green interim CP PDFs are computed on data up to the green dotted vertical lines at date $\Tilde{t}$ and mark the minimum amount of updated data that is needed to locate the global maximum of the green CP PDF less than $100$ trading days from the defined crisis onset dates. The segment fits with corresponding confidence bands are shown to guide the eye. More information can be found in the running text.}
    \label{fig: precursor discussion}
\end{figure}
In the CP analysis of the mean market correlation $\Bar{C}(t)$ from 31 January 1992 until 28 December 2012 in subsection \ref{sec:ChangePointResults}, we take a retrospective point of view to investigate whether the linear trend changes in the mean market correlation $\Bar{C}(t)$ encode information about economic crisis events of global importance. Since our findings in subsection \ref{sec:ChangePointResults} suggest indeed such a connection, in this subsection we complement the consistency of our results by changing from a retrospective to an on-line perspective. From the on-line viewpoint three key questions arise: First, is the time evolution of the CP PDFs $p(E|\Bar{C}(t),\underline{t},\mathcal{I})$ of one CP $E$ consistent with our expectations if we apply it to currently available data and update these data whenever new trading days end? Second, how much data of a changing mean market correlation trend, which potentially marks an important economic period, has to be included to render the CP analysis sensitive? And third, do the results change consistently on the stronger timely confined data segments, i.e. do the CP PDFs mirror in a way the relative impact of the events included in the sub-intervals? To answer these questions, we assume one CP and compare the CP analysis results calculated on two slightly differing types of time series segments for the dot-com bubble, the global financial crisis and the Euro crisis. The first type is limited exclusively to pre-crisis data of the mentioned major crisis events and the second type includes a small amount of mean market correlation data beyond the crises' onsets at $t_0$, i.e. it is cut after $t_1>t_0$. The results are presented in figure \ref{fig: precursor discussion}. For the pre-dot-com bubble segment as well as for the financial crisis data we used every tenth data point to avoid numerical instabilities due to an overflowing normalisation factor of the CP PDFs. The short segment of the Euro crisis is not thinned at all. Analogously to subsection \ref{sec:ChangePointResults} we show the accompanying segment fits of the orange and the black CP PDFs in figure \ref{fig: precursor discussion} to illustrate the most probable linear segments each of which might resemble specific economic periods.\\

\subsubsection{Dot-Com Bubble}
We start considering the pre-dot-com bubble data in figure \ref{fig: precursor discussion}. The blue segment is cut from the beginning of the $\Bar{C}(t)$ time series on 31 January 1992 up to the approximate onset $t_0$ of the rapid increase of the \textit{NASDAQ Composite Index} on 1 January 1998. Therefore, it does not contain any dot-com period data and the CP PDF should accumulate either around the times of economic events of major impact that are already confined in the data segment or, in the absence of such events, be less pronounced and probably located almost randomly due to some noisy outliers in the data (i.e. the probability to find the mode of the CP PDF exactly in the end period of the data is relatively low). Note that in any case we have to expect some mode of the CP PDF $p(E|\Bar{C}(t),\underline{t},\mathcal{I})$ regardless of whether there is a crisis event or not. This traces back to the design of the trend CP analysis in which we compute the CP PDF $p(E|\Bar{C}(t),\underline{t},\mathcal{I})$ conditional on the background information $\mathcal{I}$, i.e. our prior knowledge/assumption that at least one CP is present in the data. Keeping that in mind, the CP PDFs will exhibit a flat profile only in the rare situation of almost no fluctuations of constant trends, i.e. in the limiting case of an almost unperturbed straight line.\\
However, the CP PDF $p(E|\Bar{C}(t),\underline{t},\mathcal{I})$ indeed accumulates over the year 1995 coinciding with two important economic events: the so-called 1994 bond market crisis \cite{o:bond_massacre} with sharply decreasing bond prices over 1994 (and a strong drop in the yield spread of 1-year and 30-year U.S. Treasury Bonds from roughly \SI{2.83}{\percent} in 25 January 1994 to \SI{0.54}{\percent} in 13 December 2004 \cite{o:yield_data}, i.e. an indicator for a heavily flattened yield curve and worse economic conditions), and the Mexican peso crisis \cite{o:mexican_peso_crisis} starting in December 1994 which spread to other emerging markets. Updating the time series segment with new available data and based on the results of subsection \ref{sec:ChangePointResults}, we expect that the CP PDF is shifted consistently to the dot-com bubble onset period in 1999, once data beyond the onset is included. And in fact, the black CP PDF computed on the updated red mean market correlation $\Bar{C}(t)$ including data up to 1 June 1999 is shifted from 1995 to January 1999 corresponding to the dot-com bubble CP results in subsection \ref{sec:ChangePointResults}.\\
Note that the appearance of the black CP PDF's peak cannot be interpreted in terms of a precursor of the upcoming bubble for two reasons. First, we need some data points beyond the crisis onset and second, from the on-line point of view the shifted CP PDF might be in turn corrected by newly updated data points that uncover the detected trend change as a negligible fluctuation in the long run. Nevertheless, once detected, subsequent inclusion of new information can also confirm the shifted CP PDF, being the case for the major events detected in subsection \ref{sec:ChangePointResults}. In that way the on-line implementation can be very helpful to focus economic research attempts early onto current time periods which in terms of impact supersede the last major event --- here i.e. the bond crisis 1994 and the Mexican peso crisis in December 1994 --- and have potentially stronger influence to future economic states. Such a line of argument might be supported to a certain extent by the clearly more pronounced and narrow profile of the black CP PDF in Q1 1999 compared to the wider and less pronounced orange CP PDF of the years 1995, corresponding to a sharper and more probable trend change in 1999 than in 1995.\\
Additionally, we determined approximately the minimum amount of data $\Bar{C}(t)$ that are needed to shift the global maximum of the CP PDF to a date less than $\pm 100$ trading days earlier/later as the bubble's onset date defined by the thinned time series date that follows directly on the bubble's onset on 1 January 1999, i.e. the 5 January 1999. In detail this procedure leads to the 30 April 1999 marked by the green dotted vertical line in figure \ref{fig: precursor discussion}. Thus, to make the CP analysis sensitive to the bubble onset we need approximately raw stock return data up to the 1 June 1999 (i.e. $21$ trading days later), since the mean market correlation $\Bar{C}(t)$ is derived from windows of $42$ trading days returns and identified with the 21st window date.\\
The resulting green CP PDF computed on data up to the green dotted vertical line at date $\Tilde{t}$ fits into the previous reasoning as it represents a snapshot at date $\Tilde{t}$ of the time evolution from the orange CP PDF at $t_0$ to the black CP PDF at date $t_1$ with $t_0<\Tilde{t}<t_1$. The old global maximum of the orange CP PDF shrank down to the flat double peaked relic of the green PDF and the new global maximum in January 1999 has already manifested itself. By adding new data the green global maximum will eventually grow to the black CP PDF and the green remnants in 1994/1995 vanish.

\subsubsection{Financial Crisis}
For the financial crisis we observe a very similar behaviour. The segments start beyond the dot-com bubble's trough on 1 October 2002. The pre-crisis segment is truncated with the burst of the U.S. housing bubble as the seed event of the financial crisis on 1 January 2007, whereas the longer red time series includes in-crisis data until 9 August 2007 when the interest rates for inter-bank financial loans rose sharply. Similar to the previously discussed dot-com bubble segment the orange CP PDF might reflect important events in the confined context of the segment. Consistently with the coincidence of the CP PDFs and the U.S. housing bubble's burst around January 2007, the onset of the bubble's pronounced rise began in 2003 \cite{a:griffin2021}. Additionally, rising oil prices in 2003 from  \$$20$/bbl to over \$$30$/bbl peaking in 2008 with \$147/bbl, a phenomenon referred to as the 2000s energy crisis \cite{o:oil_energy_crisis, o:oil_price_data}, may have contributed to the trend change. Also the Venezuelan general strike in 2002/2003 might play a role because it hindered oil exports heavily. In this time the U.S. imports of oil from Venezuela show a pronounced trough \cite{o:venezuela_general_strike, o:USoilImports}. The ambiguity and plurality of possible events underlines the obvious limitations of such a CP analysis which is unable to distinguish the cause of the trend change when the events overlap. However, it might help to reduce the number of probable events that could be responsible for new economic states to the area of pronounce CP PDFs.\\
The green CP PDF is computed with data until 1 March 2007, i.e. the approximate date from which on the CP PDF is sensitive to the U.S. housing bubble burst is around 30 March 2007. The strongly accumulated probability mass in the narrow green global maximum is smoothed out a bit to the black CP PDF when the data is updated.

\subsubsection{Euro Crisis}
In contrast to the former analyses the Pre-Euro crisis segments tell a different story due to the fact that the crisis is embedded in turbulent economic times which make it rather difficult to isolate a precise period of the Euro crisis. In fact the mean market correlation shows alternating gradually increasing linear downwards trends interrupted by less pronounced upwards trends, i.e. an overall downwards trend in the long run. Therefore, the detected mode of the CP PDF is more sensitive to the chosen time series period. This is in good agreement to our CP analysis in retrospect (cf. subsection \ref{sec:ChangePointResults}) where the CP PDF modes in the appropriate period are also widened from 2008 until 2010. Anyway, the applicability of the CP analysis is limited due to the alternating short term trends and the non-isolated nature of the Euro crisis emerging almost directly after the global financial crisis.\\
In any case we follow our previous approach for the starting time of the investigated time series segments, i.e. we start with the culmination event of the previous crisis, here the Lehman Brothers' bankruptcy on 15 September 2008. Depending on the end of the segment the CP PDF is shifted consistently to the right, i.e. the emerging Euro crisis supersedes older events, but cannot be fixed exactly. It fluctuates depending on the amount of included in-crisis data which of course add more short term trend information and compromise the assumption of only one CP more and more. Varying the end date from our defined onset of the Euro crisis on 1 January 2009 to 1 June 2009 leads to the CP PDF fluctuating roughly between 1 January and 1 April 2009. However, the onset of the negative long term trend is slightly before 1 January 2009. The corresponding analysis of segments ending on 1 December 2008 and 1 January 2009 is presented in figure \ref{fig: precursor discussion} and underlines the missing isolation of the Euro crisis period from the financial crisis and the interconnected Great Recession of 2008 when the U.S. gross domestic product fell \SI{4.3}{\percent} which is the deepest recession since World War II \cite{o:recession, o:recession_GDP}.\\
In contrast to the former two analysed segments, the orange and black CP PDFs do not differ a lot in shape suggesting that the related trend changes cannot be classified in terms of lower or higher impact, but seem to be of similar nature constructing the downwards trend in the long run. This seems to be feasible because over 2008 the critical economic conditions led to several giant rescue plans for banks and a cascade of severe stock market crashes. From 6 to 10 October 2008 the \textit{Dow Jones Industrial Average} fell \SI{18.2}{\percent}, the \textit{S\&P500} even over \SI{20}{\percent} \cite{o:crash6to10October2008}. Some crashes created the need of interim interruption of the stock markets, e.g. in Indonesia (from 8 to 13 October 2008) due to drop of $\SI{10}{\percent}$ in one day or Iceland (9, 10 and 13 October 2008) \cite{o:Indonesia_stock_crash,o:Iceland_stock_crash}. On 24 October 2008 again losses of around \SI{10}{\percent} were realised in most of the indices \cite{10percentstocks}. This as well as the IMF prediction of a worldwide recession of \SI{-0.3}{\percent} on 6 November 2008 and accompanying lowering of target interest rates by the Bank of England and the ECB contribute most probably to the detected CP of the orange PDF \cite{o:6Nov2008}. By updating the data up to $t_1$ finally the black CP PDF emerges slightly after 1 December 2008 when the National Bureau of Economic Research officially declared that the U.S. was in a recession since December 2007. In consequence of these news the \textit{S\&P500} lost \SI{8.93}{\percent} and the financial stock of the index even \SI{17}{\percent} \cite{o:December2008Crash}. Further hints supporting the hypothesis that the event cascade drives the downwards trend is given by the green interim CP PDF result: As the data is updated, the green interim CP PDF is first widely smeared out between the old orange and the new black PDF peak and becomes a bi-modal CP PDF with the peaks near these two positions. In that way, the time evolution of the CP PDF resembles the cascade of similarly important events that contribute to the decline of the mean market correlation in the long run. In principle, also in the case of further added data the CP PDF travels along the cascade of events via a bi-modal interim PDF from peak to peak in a similar pattern.\\
The method is sensitive to the short term trend change if data until 23 December 2008, marked by the green vertical dotted line, is included. This corresponds to an earliest CP analysis sensitivity  from 26 January 2009 on.\\

\section{Discussion and Conclusion}
\label{sec:Discussion}
In sections \ref{sec:ChangePointResults} and \ref{sec:OnlineChangePointResults} we present the results of two change point analyses: first, in retrospect over the thinned mean market correlation $\Bar{C}(t)$ time series and second, in an adaptive approach that updates a pre-crisis segment continuously as present trading days end. Based on an extensive literature research we provide a detailed comparison to the overall economic history of roughly 20 years from 31 January 1992 to 28 December 2012 that include three major economic crises and check the consistency of our results against a variation of the method's intrinsic parameter, i.e. the number of expected CPs. Applying our computationally efficient implementation of the CP analysis from one up to five CPs (roughly $3\cdot 10^8$ combinations) we find a weak dependence of the main results of the CP distributions on the expected number of CPs from two up to five CPs, whereas one CP seems to be a too simple assumption over 20 years of economic evolution. Given that, we find consistent and reasonable results from sub-periods up to the full period: The retrospective analysis of the whole dataset suggests a connection between the CP PDFs of linear trend changes in the mean market correlation and major economic events as the dot-com bubble, the global financial crisis and the Euro crisis. Additionally, lower peaks seem to accumulate in crisis periods of lower impact. Naturally, the analysis is not feasible to infer on causal dependencies between trend changes of the mean market correlation and single economic events, but our studies support the idea of the mean market correlation (i.e. the market mode) to be an informative measure of macroeconomic changes as stated in the literature (cf. \cite{RandomMatrixLalouxPotters,RandomMatrixStanley,Stepanov2015}). We note that the onset of the global financial crisis is especially reflected by the CP PDF peaks of the U.S. housing bubble burst and might illustrate the strong interdependence of the bubble and the global crisis.\\
Furthermore, the CP PDFs divide the considered period into intervals that match rather well with the locally stable market state intervals found in an article of Stepanov et al. \cite{Stepanov2015} by a clustering approach: The authors in \cite{Stepanov2015} find the first four years to be a rather calm period in cluster states one and two which is mixed with some intermediate state five around Spring 1996. This is reflected by weakly pronounced CP PDF peaks at that time. A more notably transition occurs over the dot-com bubble from 1999 to 2003 into a fully intermediate transition regime dominated by states three, four, five and six. The transition is marked by the CP PDFs of the dot-com bubble and by an increased CP probability in the whole intermediate period up to 2007. Around this date the authors in Stepanov et al. define states seven and eight to be economic turbulent crisis states which is in accordance to the high CP PDFs in the end of 2006 and beginning of 2007 as well as the pronounced local CP PDF modes in the turbulent period from 2007 to the end of 2012. Interestingly, we can reconstruct approximately the economic state structure presented in Stepanov et al. by inferring them directly from the trend changes of the mean market correlation. Our independent methodological CP approach adds new evidence to the general reasoning of locally (meta)stable market states. The existence of quasi-stationary states is further supported by an independent resilience analysis on the \textit{S\&P500} mean market correlation time series which suggests locally stable economic states eventually operating on multiple time scales \cite{MemoryEffectsSP500}.\\
The on-line approach assuming one CP on the pre-crisis segments that are updated with in-crisis data over time yields further insights into the methodology's sensitivity and the relation of CP PDFs and the considered time period. First, the CP PDF is consistently shifted from left to right in times of crises' onsets for the dot-com bubble and the financial crisis. The shift becomes manifest in our analysis by including data of roughly 80 to 100 trading days after the crisis onset. The character of the Euro crisis seems to be different from the former two because of the almost direct transition from the financial crisis, i.e. missing isolation of the events in the time dimension. The CP PDF is jumping roughly between a stock market crash on 1 December 2008 and April 2009 depending on the time series length. This might indicate a cascade of non-isolated events in that period. Note that this is also reflected by the wider modes of the CP PDFs in the retrospective analysis in section \ref{sec:ChangePointResults} between 2008 and 2010. Furthermore, zooming into the pre-crisis segments instead of analysing the whole $\Bar{C}(t)$ time series changes the resolution of the detected trends in relation to the events included in the segments. For example, the highest peaks in the retrospective analysis are around the financial crisis onset and after the Euro crisis' culmination event of Greece's downgrading. More local events as the bond crisis 1994, the Mexican peso crisis or Asian and Russian financial crisis are not or less pronounced in relation to the included major events. After exclusion of the major crisis events by zooming e.g. into the pre-dot-com bubble segment, the CP PDF accumulates around the bond crisis 1994 and Mexican peso crisis. This means the CP PDFs might reflect up to a certain degree an economic impact ranking of the events relative to the most important ones that are included in the analysis segment. Moreover, such an observation implies that the zoom changes the event impact resolution of the CP analysis.\\
Naturally, the on-line approach cannot be seen as a precursor tool for economic crisis, since it needs a certain minimum amount of in-crisis data and more important, a recently detected CP might be identified as part of fluctuations if again new data is available. However, the overall results imply that the mean market correlation $\Bar{C}(t)$ contains information about changing market dynamics and global crisis events. This gives rise to the idea of designing leading indicators and precursors based on the market mean correlation $\Bar{C}(t)$. Related scientific attempts of precursor design based on correlations in general can be already found in the literature \cite{Zheng2012,Heckens_2020,Heckens_2022}.\\
Anyways, although the presented study suggests the mean market correlation to be a promising economic indicator, it is way too early for final deductions on the mean market correlation's role as economic information measure. Future research on this topic might consider other time periods applying similar methodologies as the one used here or even find different and adequate procedures for possibly even causal inference on the relation of mean market correlation and macroeconomic events.\\
Furthermore, the CP analysis by design detects almost always a CP apart from the limiting case of an unperturbed straight line. In that sense, it might be advisable to assure the detected trend change by involving further analyses. For example, the computed segment fit can be compared to a straight line via a Bayesian model comparison \cite{von2014bayesian} or information criteria \cite{Akaike1998} to infer whether the number of assumed CPs is a justified condition or not. However, due to the definition the information criteria might always prefer the straight lines for time series that include only a few data beyond the CP. Assigning higher weight to the most recent data might be a possible way to remedy this problem.

\section*{Data and Software Availability}
The simulated data and Python codes are available on github via \url{abc} under a \textit{GNU General Public License v3.0}. The open source python-implementation is named \textit{antiCPy} and can be found at \url{https://github.com/MartinHessler/antiCPy} under a \textit{GNU General Public License v3.0}. The raw data can be downloaded via \cite{yfinance} and the preprocessed time series is available via \cite{DataZenodo}.

\section*{Author Contributions}
Conceptualization, M.H., T.W. and O.K.; methodology/software, M.H.; validation/interpretation, M.H. and T.W.; formal analysis/investigation, M.H.; data resources, T.W.; data curation, M.H. and T.W.; writing --- original draft preparation, M.H.; writing --- review and editing, M.H., T.H. and O.K.; visualization, M.H.; supervision, O.K. All authors have read and agreed to the published version of the manuscript.

\section*{Additional Information}
This research received no external funding and the authors declare no conflict of interest.

\section*{Acknowledgments}
M.H. and T.W. thank the Studienstiftung des deutschen Volkes for a scholarship including financial support.
\newpage
\section*{Abbreviations}
The following abbreviations are used in this manuscript:\\

\noindent 
\begin{tabular}{@{}ll}
CB & Credibility Band \\
CP & Change Point\\
ECB & European Central Bank\\
FED & U.S. Federal Reserve System \\
NASDAQ & National Association of Securities Dealers Automated Quotations\\
U.S. & United States \\
USA & United States of America \\
PDF & probability density function\\
Q1, Q2, ... & quarter 1, quarter 2, ...\\
S\&P & Standard and Poor's
\end{tabular}

\section*{Appendix}
\appendix
\setcounter{figure}{0}
\renewcommand\thefigure{A.\arabic{figure}}
\section{Detailed Data Preparation}
\label{Appendix A}
Daily stock data from the 500 companies that are included in the \textit{S\&P500} stock index which is traded at U.S. stock exchanges, were downloaded via the Python package \textit{yfinance} \cite{yfinance}. Confined to the time period between 1 January 1992 and 28 December 2012 which is chosen analogously to \cite{Rinn2015DynamicsOQ}, we filter for companies that are included in the \textit{S\&P500} index for at least $\SI{99.5}{\percent}$ of the time and linearly interpolate occasionally missing price data $P_t$ via the function \textit{interpolate()} of the Python package \textit{pandas}. We end up with 249 daily-resolved stock price time series $P_t$ for 5291 trading days. Analogously to \cite{SCHAFER20103856} the returns $R_t = (P_{t+1} - P_t)/P_t$ are locally normalised as
\begin{equation}
    r_t = \frac{R_t - \langle R_t\rangle_n}{\sqrt{\langle R^2_t\rangle_n - \langle R_t\rangle^2_n}}
\end{equation}
with the local mean $\langle\dots\rangle_n$ across the $n=13$ (cf. \cite{Mnnix2012}) most recent data points to
remedy the impact of sudden changes in the drift of the time series. The procedure corresponds to a local standard normalisation transformation with the local mean $\langle R_t\rangle_n$ and standard deviation $\sqrt{\langle R^2_t\rangle_n - \langle R_t\rangle^2_n}$. 

\begin{figure}[h!]
    \centering
    \includegraphics[width = \textwidth]{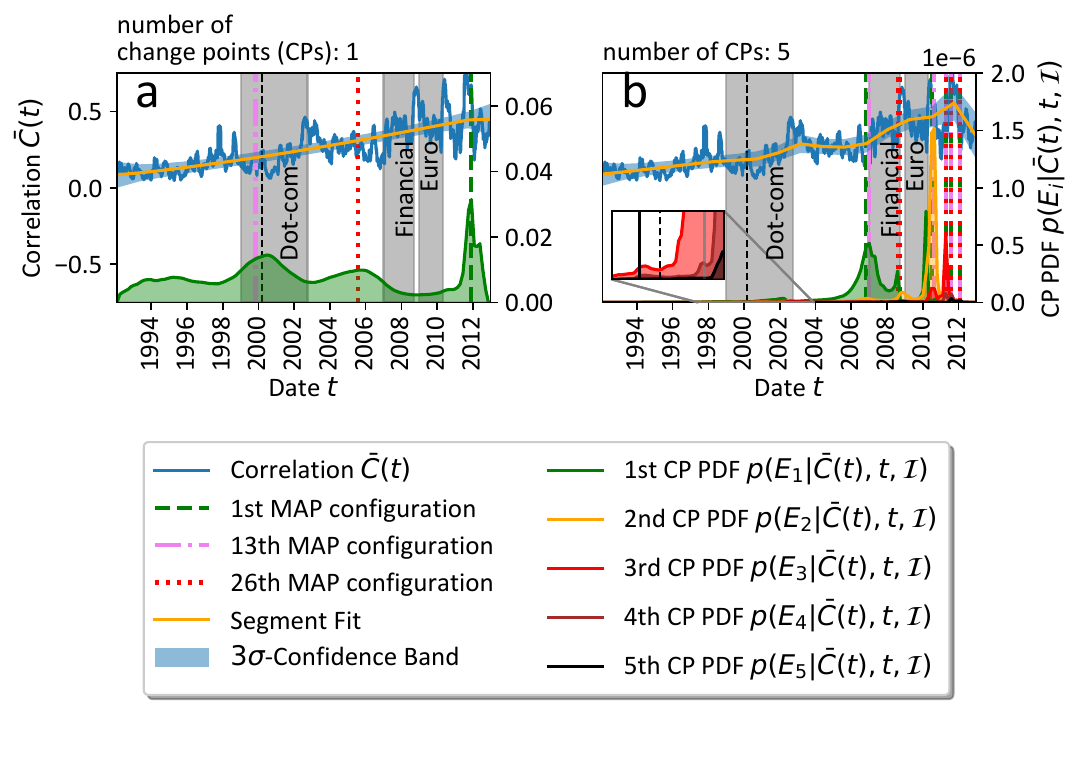}
    \caption{Additional results of the CP analysis performed in the main article. (a,b) CP analysis for one and five CPs. The distributions are in good agreement with the results for two up to four CPs and underline the consistency of the results.}
    \label{fig:SI CP}
\end{figure}

Local correlation coefficients 
\begin{equation}
\label{eq:Correlation}
    C_{i,j} = \frac{\langle r^{(i)}_t r^{(j)}_t\rangle_\tau - \langle r^{(i)}_t \rangle_\tau  \langle r^{(j)}_t \rangle_\tau}{ \sigma^{(i)}_\tau \sigma^{(j)}_\tau  }
\end{equation}
are calculated over a time period of $\tau=42$ trading days (cf. \cite{Mnnix2012}) with the local standard deviations $\sigma^{(i)}_\tau$. Shown by the principle component analysis in \cite{Stepanov2015}, the mean correlation
\begin{equation}
    \Bar{C} = \frac{1}{N} \sum_\textmd{i,j} C_{i,j}
\end{equation}
is in high agreement with the first principle component. Furthermore, it describes the greatest part of the variability in the data. Thus, we focus on the mean market correlation $\Bar{C}(t)$ with each mean correlation calculated over 42 trading days and identified with the 21st value of each data window.

\section{Advanced Change Point Analysis}
\label{Appendix B}
The CP analysis results are complemented by the computations under the assumption of one and five CPs. In figure \ref{fig:SI CP} the corresponding PDFs (a,b) are provided. The overall similarities between the PDF results from one up to five CPs underline the consistency of the results. To complete the advanced analysis we performed a search for the earliest occurrence of a CP in the dot-com bubble period: for the assumption of one and five CPs the seventh and 852th most probable configuration, respectively, includes the dot-com period for the first time. This corresponds to the most probable $\SI{5.0}{\percent}$ and $\SI{3.0}{\cdot10^{-4}\percent}$ of possible CP configurations for the assumption of one or five CPs, respectively.

\FloatBarrier

\bibliography{references}

\end{document}